\documentclass{aastex631}
\turnoffeditone

\usepackage{multirow}
\usepackage[T1]{fontenc}

\shorttitle{ILT imaging of M 51 and SN 2011dh}
\shortauthors{Venkattu et al.}

\graphicspath{{./}{figures/}}

\begin{document}

\title{Sub-arcsecond resolution imaging of M 51 with the International LOFAR Telescope  \footnote{Submitted on March 31, 2023; Accepted on June 28, 2023}}

\author[0000-0001-9896-6994]{Deepika Venkattu}
\affiliation{The Oskar Klein Centre, Department of Astronomy, Stockholm University, AlbaNova, SE-10691 Stockholm, Sweden \\}
\correspondingauthor{Deepika Venkattu}
\email{deepika.venkattu@astro.su.se}

\author[0000-0002-3664-8082]{Peter Lundqvist}
\affiliation{The Oskar Klein Centre, Department of Astronomy, Stockholm University, AlbaNova, SE-10691 Stockholm, Sweden \\}

\author[0000-0001-5654-0266]{Miguel Pérez Torres}
\affiliation{Instituto de Astrofísica de Andalucía, Glorieta de la Astronomía, s/n, E-18008 Granada, Spain \\
}
\affiliation{Facultad de Ciencias, Universidad de Zaragoza, Pedro Cerbuna 12, E-50009 Zaragoza, Spain \\}

\author[0000-0003-0487-6651]{Leah Morabito }
\affiliation{Centre for Extragalactic Astronomy, Department of Physics, Durham University, Durham DH1 3LE, UK\\}
\affiliation{Institute for Computational Cosmology, Department of Physics, University of Durham, South Road, Durham DH1 3LE, UK\\ }

\author[0000-0002-8079-7608]{Javier Moldón}
\affiliation{Instituto de Astrofísica de Andalucía, Glorieta de la Astronomía, s/n, E-18008 Granada, Spain \\
}
\author[0000-0003-2448-9181]{John Conway}
\affiliation{Department of Space, Earth and Environment, Chalmers University of Technology, Onsala Space Observatory, 439 92 Onsala, Sweden \\}

\author[0000-0002-0844-6563]{Poonam Chandra}
\affiliation{National Radio Astronomy Observatory, 520 Edgemont Road, Charlottesville, VA 22903, USA \\}

\author{Cyril Tasse }
\affiliation{GEPI \& ORN, Observatoire de Paris, Université PSL, CNRS, 5 Place Jules Janssen, 92190 Meudon, France and Department of Physics \& Electronics, Rhodes University, PO Box 94, Grahamstown, 6140, South Africa\\}

\begin{abstract}
We present an International LOFAR Telescope sub-arcsecond resolution image of the nearby galaxy M 51 with a beam size of \SI{0.436}{\arcsecond} $\times$ \SI{0.366}{\arcsecond} and rms of 46 $\mu$Jy. We compare this image with an European VLBI Network study of M 51, and discuss the supernovae in this galaxy, which have not yet been probed at these low radio frequencies. We find a flux density of 0.97 mJy for SN 2011dh in the ILT image, which is about five times smaller than the flux density reported by the LOFAR Two-metre Sky Survey at \SI{6}{\arcsecond} resolution using the same dataset without the international stations. This difference makes evident the need for LOFAR international baselines to reliably obtain flux density measurements of compact objects in nearby galaxies. Our LOFAR flux density measurement of SN 2011dh directly translates into fitting the radio light curves for the supernova and constraining mass-loss rates of progenitor star. We do not detect two other supernovae in the same galaxy, SN 1994I and SN 2005cs, and our observations place limits on the evolution of both supernovae at radio wavelengths. We also discuss the radio emission from the centre of M 51, in which we detect the Active Galactic Nucleus and other parts of the nuclear emission in the galaxy, and a possible detection of Component N. We discuss a few other sources, including the detection of a High mass X-ray Binary not detected by LoTSS, but with a flux density in the ILT image that matches well with higher frequency catalogues. 
\end{abstract}

\keywords{radio continuum: galaxies - galaxies: individual (M 51) - techniques: high angular resolution - supernovae: individual (SN 2011dh) - supernovae: individual (SN 1994I) - supernovae: individual (SN 2005cs) - AGN - HMXB}

\section{Introduction} \label{sec:intro}

The low frequency radio sky is being explored extensively with instruments like the LOw Frequency ARray (LOFAR; \citealt{2013A&A...556A...2V}), the Giant Metrewave Radio Telescope (GMRT; \citealt{2017A&A...598A..78I}) and the Very Large Array (VLA) Low Band Ionospheric and Transient Experiment (VLITE; \citealt{2016ApJ...832...60P}). Along with the onset of the Square Kilometre Array (SKA) era, there has never been a stronger focus on lower frequencies in the radio band. The most recent addition to this discovery space is the tested framework for sub-arcsecond resolution imaging with the International LOFAR Telescope (ILT; \citealt{2022A&A...658A...1M}). ILT observations have proven to be very useful to unveil the nature of the radio emitting sources in compact regions of nearby luminous infrared galaxies (e.g., detection of steep spectrum outflows: \citealt{2016A&A...593A..86V}, \citealt{2018A&A...610L..18R}; probing thermal absorption in their nuclei: \citealt{2022A&A...658A...4R}), the complex structures in the radio lobes of Active Galactic Nuclei (AGNs; \citealt{2022A&A...658A...5T}), the life cycle of radio galaxies \citep{2022A&A...658A...6K} and much more. In this work, we present the first sub-arcsecond resolution image of the nearby galaxy M 51 which resulted in the detection of 12 sources at 145 MHz.

The Whirlpool galaxy M 51a (NGC 5194) is an almost face-on spiral galaxy with an interacting smaller companion M 51b (NGC 5195). At a distance of 8.4 Mpc \citep{2012A&A...540A..93V}, it has been extensively studied at various wavelengths and is a very good candidate for population studies. For example, \cite{2015MNRAS.452...32R} present a radio survey of M 51 with the European Very Long Baseline Interferometry Network (EVN) and discuss star-formation, supernovae (SNe) and nuclear emission. \cite{2007AJ....133.2559M} study the compact sources in M 51 with the VLA, complemented by multiwavelength data in the optical and X-ray. Radio emission remains unattenuated by dust and is hence an extinction-free tracer of star formation \citep{1992ARA&A..30..575C}. In addition, low-frequency radio observations are known to be primarily composed of synchrotron radiation with less than 10 \% of continuum emission from thermal radiation \citep{2017ApJ...836..185T}.

In this context, a study of a nearby galaxy like M 51 with the ILT is especially interesting because of the instrument's high angular resolution. With the inclusion of the international stations, the effective area of the beam is about a factor 200 smaller than the beam area of standard LOFAR surveys with the Dutch stations only. This huge improvement in resolution implies that compact sources can be more easily disentangled, and that ILT observations are essentially free from contamination from diffuse flux density. 

Probing high brightness compact sources within nearby galaxies, like supernovae, supernova remnants, AGNs, etc., with the ILT remains unexplored and could even lead to detections of new classes of compact objects. This study of M 51 furnishes a proof of concept of this new discovery space. In the future, with automated processing of ILT data of nearby galaxies, large populations of compact sources could be probed. For objects like radio supernovae with longer emission periods at lower frequencies, along with the large sky coverage of LOFAR, this could mean a potential survey of all historical supernovae in the galaxies surveyed by the ILT in the Northern hemisphere in the near future. Since radio luminosity of a supernova also correlates with higher progenitor wind density, the 150 MHz radio emission could help constrain different supernovae types with different progenitor stars from the mass loss rates obtained.

We focus mostly on the low-frequency radio emission of the historic supernovae detected in M 51 as our chief interest, but also discuss briefly the detection of the nucleus of M 51, as this is the highest angular resolution observation of M 51 in this frequency range. We emphasize again the aspect of the beam area provided by the ILT observations, which is similar to, or even better than, the resolution provided by world-class radio interferometers working at GHz frequencies, such as the VLA. This allows us to combine existing VLA observations (see Table \ref{tab:2005csLog}) with ILT observations, and determine the radio spectral energy distribution and the light curve of these SNe in a meaningful way. This will involve a much greater degree of uncertainty if lower resolution survey observations are used, as we also discuss in Section \ref{sec:results}. 

The low frequency radio emission from SNe is known to arise at later times than at high frequencies, i.e., when the blast wave has propagated further away from the progenitor. It therefore probes mass loss at earlier times before the explosion, allowing for instance, the detection of possible changes in mass loss prior to explosion. For radio SNe which have been studied extensively at higher radio frequencies, observing low frequency spectral turnovers can help distinguish absorption models like free-free absorption (FFA) and synchrotron self-absorption (SSA) providing additional constraints on magnetic field strength, and wind properties of density and temperature.

In Section \ref{sec:obs} we present the ILT observations and describe briefly the data reduction process for this. In Section \ref{sec:results} we present the the sources detected and discuss our results with respect to previous work, with a focus on supernovae and the nuclear emission in M 51. Finally Section \ref{sec:conc} presents the conclusions. 

\section{Observations} \label{sec:obs}
\subsection{LOFAR observations}

LOFAR data are obtained from project LC2\_038 (PI: H.Rottgering) observed on 10th September 2014, as a part of the LOFAR Two-Metre Sky Survey (LoTSS, \citealt{2019A&A...622A...1S}). The second LoTSS data release (LoTSS-DR2; \citealt{2022A&A...659A...1S}) provides a mosaic of the pointing P39Hetdex19 from this project, with M 51 close to the centre. 
We successfully obtained an ILT image of the galaxy using the same data, which includes international stations but which have not been included in the survey dataproduct. This resulted in the first sub-arcsecond VLBI image of the galaxy M 51 and a supernova in it at frequencies as low as 145 MHz. 
The observation using the High Band Antenna (HBA; 120-240 MHz) follows the standard strategy for LoTSS , which is a 10 minute observation of a bright flux density calibrator (3C 196 in this case) before and after the 8 hour on-source observation. Since this observation is from observation cycle 2 in 2014, only eight international stations are included (five in Germany (Effelsberg, Unterweilenbach, Tautenburg, Jülich and Potsdam), and one each in the UK (Chilbolton), France (Nançay) and Sweden (Onsala)). This is unlike the present day where LOFAR routinely observes with 14 international stations, with more new stations being planned. The longest baseline length for this work is therefore 1297 km between the stations in France and Sweden (which yields an angular resolution of \SI{0.329}{\arcsecond} at 145 MHz). The observation bandwidth is the standard surveys frequency range of 120–168 MHz that avoids Radio Frequency Interference (RFI) above this frequency range. The data were recorded with 2 second sampling and in channels of 24.41 kHz width (8 channels per subband, 244 subbands).

The first part of the data reduction involves the use of PREFACTOR (\url{https://github.com/lofar-astron/prefactor}; \citealt{2019A&A...622A...1S}) for direction-independent calibration of the Dutch stations (core and remote). A model with suitably high resolution was used for 3C196\footnote{3C 196 model courtesy of A. Offringa} in the PREFACTOR pipeline with CS001 as the reference station. All stations produced good solutions for Total Electron Content (TEC) and bandpass for both the calibrator and target.

The data were analysed with v4.0.0 of the LOFAR-VLBI pipeline (\citealt{2022A&A...658A...1M} introduces the pipeline and describes v3.0.0). The latest version of the pipeline uses self-calibration optimised for LOFAR data \citep{2021A&A...651A.115V}, using NDPPP \citep{2018ascl.soft04003V} and WSClean \citep{2014MNRAS.444..606O}. The pipeline uses the LoTSS and Long Baseline Calibrator Survey (LBCS; \citealt{2022A&A...658A...2J,2016A&A...595A..86J}) catalogue servers for useful information about the sources in our target field and for nearby phase calibrators respectively. The LBCS source L332164 (ILTJ132703.36+470543.5), 0.50803 degrees away, was used for an initial in-field correction of the phases and complex gains.  

The pipeline forms a single \edit1{super station} ST001 combining all the core stations, which have already been phase-corrected using the PREFACTOR solutions. Unlike the pipeline described by \cite{2022A&A...658A...1M}, in the latest version, the delay calibration part of the pipeline is done by the facet self-calibration script in the form of three types of solutions done with the delay calibrator, which are then transferred onto the nearby target source (scalarphasediff, scalarphase and scalarcomplexgain). The end products are a self-calibrated image of the delay calibrator along with an h5parm file of solutions to be transferred. These solutions are applied by the next part of the pipeline which takes this solution-applied dataset, phase shifts it to the target's co-ordinates and splits off a smaller dataset. Further self-calibration can be carried out if necessary. Since the target imaged here was the entire galaxy M 51, in the last part of the pipeline, self-calibration of the target was not done. As a post-pipeline step, the reduced dataset now centered on M 51 was imaged using \textsc{WSClean} (\citealt{2014MNRAS.444..606O}; \citealt{2017MNRAS.471..301O}) with briggs weighting parameters resulting in an image with a beam size of \SI{0.436}{\arcsecond} $\times$ \SI{0.366}{\arcsecond} and rms of 45.85 $\mu$Jy.  

After imaging the galaxy this way using the reduced dataset from the pipeline that includes \edit1{the super station} ST001, imaging was also performed using a modified dataset which included the core stations instead of the combined \edit1{super station}, with briggs weighting parameters and a minimum uv-cut of 5 kilolambda in order to filter out some of the diffuse emission and mimic the presence of the \edit1{super station}. This resulted in an image with a much lower noise (24.05 $\mu$Jy) but a larger beam size (\SI{0.561}{\arcsecond}, \SI{0.475}{\arcsecond}), hence lower resolution and significantly more negative bowl effects than the image with the \edit1{super station}. Hence, throughout this paper, we only use the higher resolution image \edit1{with ST001} for all purposes including source detection and flux estimation.

\subsection{Source Detection}
\begin{figure}[ht!]
\includegraphics[width=\textwidth]{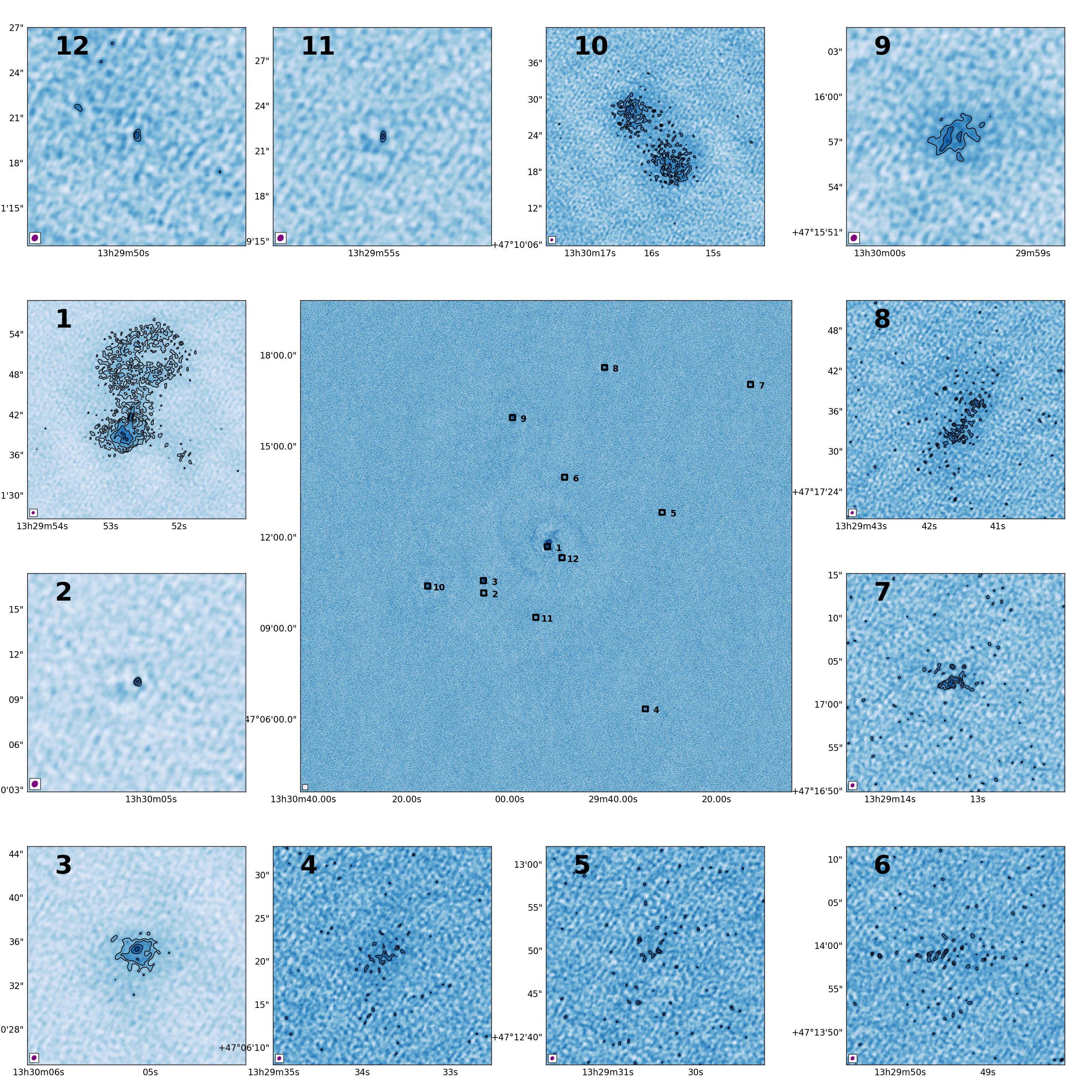}
\caption{ILT map of M 51 in the centre with sources detected using \textsc{PyBDSF} marked in the same order as in Table \ref{tab:sources}. Panels on the side show the individual sources in greater detail. Sources of note are the nuclear emission in the centre (1 is the position of the AGN) and SN 2011dh (2). \label{fig:M 51}}
\end{figure}

Figure \ref{fig:M 51} (centre panel) shows the ILT image of M 51. Source detection was performed on this image using \textsc{PyBDSF} \citep{2015ascl.soft02007M}. To enable extended sources to be detected (especially in the centre of M 51a), we set the control \textsc{PyBDSF}  parameters \verb|flag_maxsize_bm=100|, and \verb|rms_map=False|, while maintaining the threshold for island detection to be 6$\sigma$. A more detailed discussion on detecting extended sources is given in the \textsc{PyBDSF} documentation  (\url{https://pybdsf.readthedocs.io/en/latest/index.html}).  In total 12 distinct sources were identified in the ILT image by \textsc{PyBDSF} (see Table \ref{tab:sources} and the side panels of Figure \ref{fig:M 51}). Two of these sources are identified by \textsc{PyBDSF} as having subcomponents, with properties as listed in Table \ref{tab:sources} (sources 8 and 10 listed with their subcomponents). In order to compare the number of source detections and flux densities between our sub-arcsecond resolution ILT image and in the \SI{6}{\arcsecond} resolution LoTSS survey we searched for cross-identified sources in the LoTSS catalogue that lay within a \SI{2.5}{\arcsecond} radius of each LOFAR-VLBI source, resulting in 10 associated LoTSS detections. One source (source 11 with name J132954+470922 in Table \ref{tab:sources}), although detected above 6$\sigma$ in the ILT image, is not detected in the LoTSS catalogue. \textsc{PyBDSF} also predicts possible sources in the spiral arms of the galaxy but since this area has only extended emission and no cross-matches, we do not take these into account any further. We also compare our \textsc{PyBDSF} catalogue with the SNR candidates discussed in \cite{2021ApJ...908...80W}, especially the 16 radio sources that are also detected by \cite{2007AJ....133.2559M}. \edit1{Of} these, we have a possible cross-match for only one, source 17 in \cite{2007AJ....133.2559M}, and the last source in Table \ref{tab:sources} which is also not detected in the LoTSS catalogue.

\begin{table*}
\caption{Sources detected in M 51 with the ILT at 145 MHz}
\scalebox{0.9}{
\setlength{\tabcolsep}{3pt}
\renewcommand{\arraystretch}{1.5}
\hspace*{-3cm}
\begin{tabular}{lcccccccc}
\tableline\tableline
 & & \multicolumn{2}{c}{Position (J2000.0)} &  & \multicolumn{2}{c}{\textsc{PyBDSF} size} & \multicolumn{2}{c}{Cross-identification}\\
No. & Source/  & RA & Dec  & Flux &  Major & Minor & LoTSS & NED\\ 
 & Sub-component name & 13h & 47d & (mJy) & (arcsec) & (arcsec) & &\\
\tableline
1. & J132952+471142 & 29m52.62s & 11m42.83s & $0.46\pm0.3$  & $1.32\pm0.86$ &  $0.87\pm0.5$ & ILTJ132952.71+471143.3 & MESSIER 051a (centre)\\
2. & J133005+471010 & 30m05.09s  &  10m10.17s & $0.97\pm0.14$    & $0.699\pm0.08$ & $0.479\pm0.04$  & ILTJ133005.07+471011.0 & SN 2011dh\\
3. & J133005+471035 & 30m05.12s & 10m34.77s  & $71.91\pm2.5$ & $3.31\pm0.19$ & $2.8\pm0.16$ & ILTJ133005.03+471035.2 & 2CXO J133005.0+471035\\
4. & J132933+470620 & 29m33.76s & 06m20.70s & 12.71$\pm$1.73 & 6.95$\pm$0.95 & 4.12$\pm$0.56 & ILTJ132933.78+470621.3 & WISEA J132933.76+470621.4\\
5. & J132930+471250 & 29m30.45s & 12m49.50s & 12.95$\pm$2.17 & 7.56$\pm$1.27 & 5.94$\pm$0.99 & ILTJ132930.46+471250.8 & SSTSL2 J132930.49+471250.5\\
6. & J132949+471358 & 29m49.36s & 13m58.90s & 11.34$\pm$2.33 & 5.97$\pm$1.78 & 2.78$\pm$0.81 & ILTJ132949.59+471359.2 & MESSIER 051:[MCK2007] 014\\ 
7. & J132913+471702 & 29m13.24s  &  17m02.50s & 10.39$\pm$1.45 & 2.56$\pm$0.46 & 1.34$\pm$0.23 & ILTJ132913.38+471703.9 & ILT J132913.3+471704.7\\
8a. & J132941+471736 & 29m41.38s &   17m36.70s & 20.17$\pm$2.17  & 3.39$\pm$0.66 & 2.83$\pm$0.54 & ILTJ132941.60+471736.2 & SSTSL2 J132941.50+471734.9\\
8b. & J132941+471732 & 29m41.60s &   17m32.40s & 16.16$\pm$1.56  & 5.74$\pm$0.55 & 4.04$\pm$0.39 & " & "\\
9. & J132959+471557 & 29m59.50s  &  15m57.40s & 56.44$\pm$2.46 & 2.94$\pm$0.24 & 2.14$\pm$0.17 & ILTJ132959.89+471549.1 & MESSIER 051b\\
10a. & J133016+471027 & 30m16.30s  &  10m27.70s & 60.62$\pm$2.55 & 8.82$\pm$0.37 & 6.96$\pm$0.29 & ILTJ133015.94+471023.9 & 2CXO J133016.0+471024\\
10b.& J133015+471019 & 30m15.68s  &  10m19.40s & 81.56$\pm$2.97 & 10.28$\pm$0.37 & 8.13$\pm$0.3 & " & "\\%
11. & J132954+470922 & 29m54.94s & 09m22.00s & 0.83$\pm$0.17 & 0.91$\pm$0.16 & 0.49$\pm$0.06 & - & CXOU J132954.9+470922 \\
12. & J132949+471119 & 29m 49.91s & 11m 19.85s & 0.95$\pm$0.20 & 0.90$\pm$0.17 & 0.60$\pm$0.09 & - & 2CXO J132949.9+471120\\
\tableline
\end{tabular}
}
\tablecomments{Image RMS is 46 $\mu$Jy beam$^{-1}$. The same value applies for localised island rms detected by \textsc{PyBDSF} for fitting gaussians.The errors in the flux density column is the error from \textsc{PyBDSF} fitting.}
\label{tab:sources}
\end{table*}

\section{Results} \label{sec:results}

\subsection{Supernovae} 
Currently, four supernovae have been detected in the M 51 system, Type Ia SN 1945A \citep{1971AJ.....76..756K}, Type Ib/c SN 1994I \citep{1994IAUC.5961....1P}, Type IIP SN 2005cs \citep{2005IAUC.8553....1K} and Type IIb SN 2011dh \citep{2011CBET.2736....1G}, of which radio emission has been detected for SN 1994I (e.g. \citealt{2011ApJ...740...79W}) and SN 2011dh (e.g. \citealt{2012ApJ...750L..40K} and \citealt{2013MNRAS.436.1258H}). 
Only one SN is detected with LOFAR (SN2011dh is visible in both LoTSS and our ILT image), and in this section we discuss SN 2011dh and also constrain the radio emission from the other core-collapse SNe with upper limits from the ILT image which in turn constrains the supernova-circumstellar matter interaction properties.

\subsubsection{SN 2011dh}
\subsubsection{SN 2011dh ancillary data}
\begin{figure}[ht!]
\includegraphics[width=\textwidth]{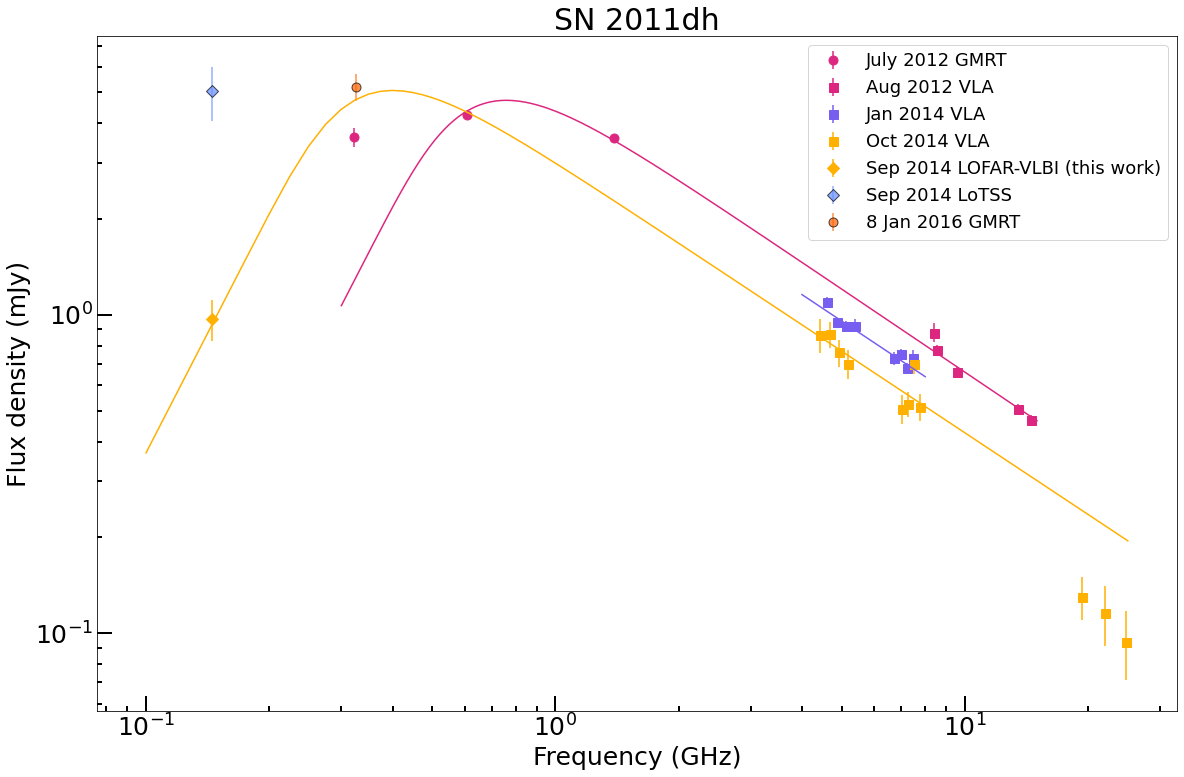}
\caption{Radio spectra of SN 2011dh at three different distinct epochs using the VLA, GMRT and ILT data in Table \ref{tab:2005csLog}. \edit1{Same colour is used to denote data considered as one epoch and included in a single fit, for e.g., September 2014 LOFAR-VLBI and October 2014 VLA observations of SN 2011dh. Data not included in the analysis has translucent markers, for e.g., data from the LoTSS catalogue at \SI{6}{\arcsecond} resolution is plotted to show the difference in flux densities. See text and Fig. \ref{fig:2011dh} for further details. } }
\label{fig:fluxplot}
\end{figure}

SN 2011dh is detected in both the LoTSS and ILT images. The ILT observation is discussed in Section \ref{sec:obs}, and the \textsc{PyBDSF} fit for SN 2011dh from the ILT image can be found in Table \ref{tab:sources} (Source number 2). Both the LoTSS and ILT flux densities for the supernova is shown in Figure \ref{fig:fluxplot}. To complement the LOFAR data of SN 2011dh, we consider other late-time radio observations of the supernova here. \\

\textbf{GMRT observations}: GMRT data from July 2012 published in \cite{2016MNRAS.459..595Y} is used in this work to complement the VLA data at a similar epoch from August 2012. \cite{2021MNRAS.507.4734R} study seven nearby galaxies with the GMRT, including M 51. The flux for SN 2011dh, clearly visible in the radio map of the galaxy, is also used in this work (private communication, Subhashish Roy). This gives an idea of the supernova at a much later epoch at lower frequencies.  \\

\textbf{VLA observations}: VLA data used for this work were from VLA programs 12A-286, 13A-370 and 14B-479 (data taken on 1 Aug 2012, 31 Jan 2014 and 18 Oct 2014 respectively\edit1{; \citealt{2019ApJ...875...17K}}). \edit1{We re-analyzed the VLA radio data for SN 2011dh on 31 Jan 2014 and 18 Oct 2014 presented and used in \cite{2019ApJ...875...17K}. For these two epochs, we find a notably higher flux, around a factor of 5 to 15, and more in line with the results of \cite{2016MNRAS.455..511D}. We suspect that there could have been a problem with the CASA version 4.3.1 pipeline used in \cite{2019ApJ...875...17K}. With these new estimates of flux densities on 31 Jan 2014 and 18 Oct 2014, the results of \cite{2019ApJ...875...17K} are valid for up to around 500 days after the explosion of the SN. }

\begin{figure}[ht!]
\includegraphics[width=\textwidth]{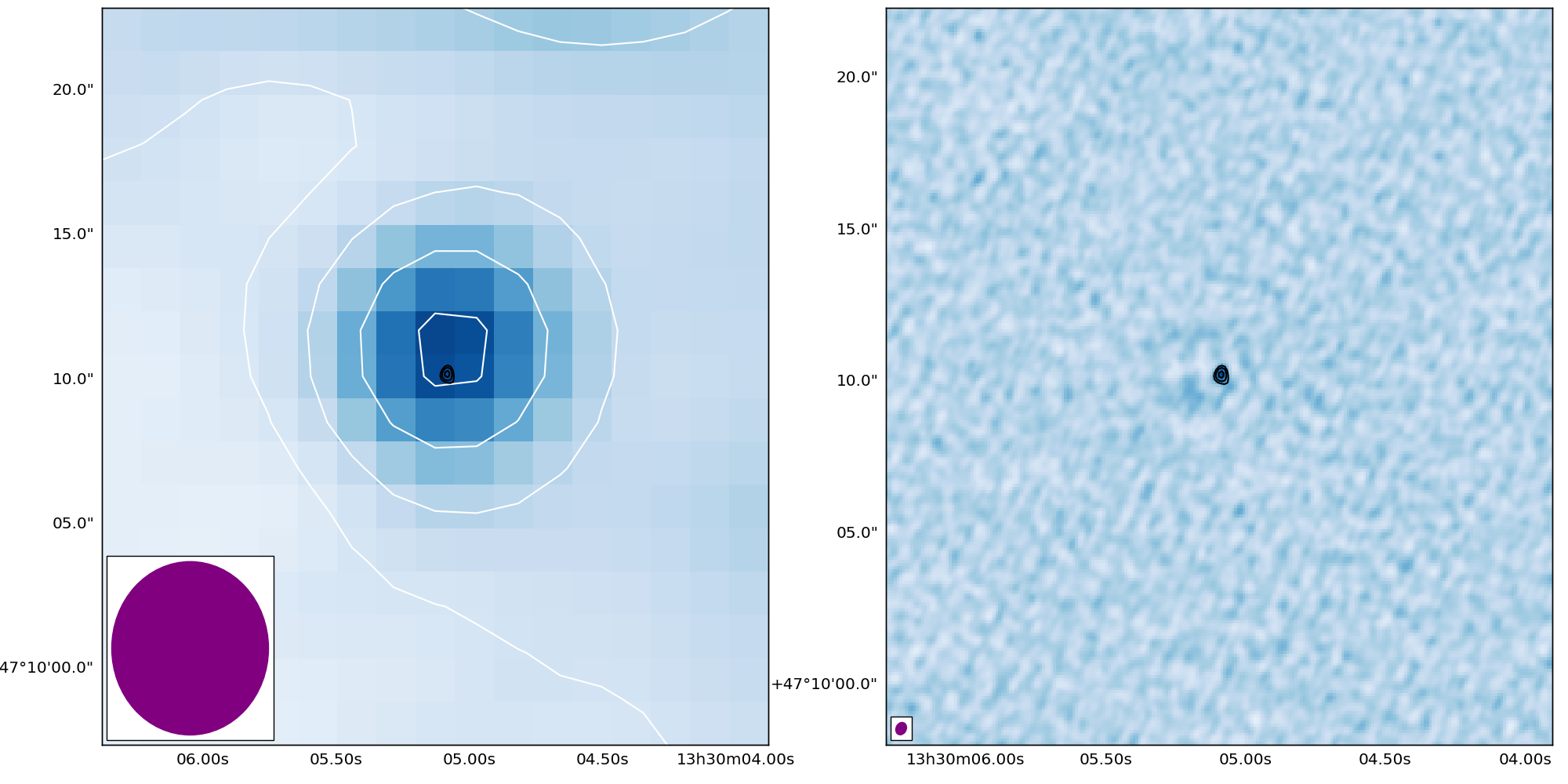}
\caption{Left: ILT image in black contours overlaid on LoTSS DR2 image with white contours. Contours from 4 $\sigma$ to 5 $\sigma$ for the ILT, and contours start from 1 $\sigma$ to 5 $\sigma$ for LoTSS. Right: ILT image in background and with black contours overlaid. Linear stretch from -5 to +5 $\sigma$. The restoring beam for the two LOFAR images are shown in the bottom left corner (LoTSS - \SI{6}{\arcsecond}; ILT - \SI{0.44}{\arcsecond} $\times$ \SI{0.37}{\arcsecond}). \label{fig:2011dh}}
\end{figure}

\begin{table*}
\caption{SSA Model Fits - SN 2011dh}
\begin{tabular}{lccccc}
\tableline\tableline
Day & $S_{\nu_{\tau}}$ & $\nu_{\tau}$ & $R_s$ & $B_s$ & $A_*$ \\

 & (mJy) & (GHz) & ($10^{16}$ cm) & (G) & \\
\tableline

497 & 4.53$\pm$0.06 & 0.64$\pm$0.01 & 4.04$\pm$0.29 & 0.08$\pm$0.002 & 9.6$\pm$0.5 \\
1197 & 4.75$\pm$1.77 & 0.34$\pm$0.05 & 7.9$\pm$1.7 & 0.04$\pm$0.01 & 22$\pm$6 \\
\tableline
\end{tabular}
\tablecomments{Parameters estimated from an SSA fit to the observed radio spectrum of SN 2011dh (Source 2 in Table \ref{tab:sources}). In the fits we have assumed a filling factor $f=0.5$ and $\epsilon_B = \epsilon _e = 0.1$. See text for further details.}
\label{tab:fitSSA}
\end{table*}

\subsubsection{SN 2011dh modelling}
SN 2011dh is a Type IIb supernova first detected by \cite{2011CBET.2736....1G}. Figure \ref{fig:fluxplot} shows the light curve with the the 2012 epoch in magenta, the January 2014 epoch in orange and the September 2014 epoch in yellow. VLA data is denoted by a square, GMRT data by a circle and LOFAR data by a diamond. The LoTSS data point (5.2 $\pm$ 0.9 mJy) is also plotted for comparison with the ILT data point, showing a large discrepancy in flux densities with the LoTSS value more than five times higher. This arises from the larger beam size of LoTSS that does not include the international stations. The higher resolution observation with a lower beam size is not as contaminated by diffuse emission from the surrounding and this translates directly into estimating parameters such as mass-loss rates of the progenitor of SN 2011dh. We note here that the LoTSS catalogue has a \textsc{PyBDSF} fit for SN 2011dh that is bigger than its beam size of \SI{6}{\arcsecond} (\SI{8.05}{\arcsecond} $\times$ \SI{7.51}{\arcsecond}), and shows a peak flux of 3.1 $\pm$ 0.4 mJy. It is likely then that the bias due to the diffuse flux from the background would be lower when considering the peak flux, rather than the total flux from the LoTSS catalogue fit much larger than the beam. 
There is also a GMRT flux density value from 2016. The curves are fit to the data using \edit1{\textsc{emcee}} \citep{2013PASP..125..306F} with an SSA (synchrotron self-absorption) model:

\begin{equation}
  \mathrm{S(\nu) = 1.582\ S_{\nu_{\tau}}  \left(\frac{\nu}{\nu_{\tau}}\right)^{5/2} \left\{ 1 - exp \left[ - \left(\frac{\nu}{\nu_{\tau}}\right)^{-(p+4)/2} \right] \right\}},  
\end{equation}
where $S_{\nu_{\tau}}$ is the flux density for synchrotron self-absorption optical depth unity, and $\nu_{\tau}$ is the frequency at which this occurs. For this model \citep{2012ApJ...750L..40K} find a value of p=2.8, and an observed peak radio flux density at $\nu_{op}=1.17\nu_{\tau}$.
\edit1{We find a value of p=2.74 and p=2.71} and observed peak radio flux density occurring at $\nu_{op}=1.178\nu_{\tau}$ and  $\nu_{op}=1.184\nu_{\tau}$ for the 2012 and 2014 epochs, respectively. Values for the fit parameters are shown in Table \ref{tab:fitSSA}. 
 
The synchrotron emission is assumed to arise as a result of the interaction of the supernova ejecta with circumstellar matter, and the fit parameters can be used to estimate some other important parameters, namely

\begin{equation}
  \mathrm{R_s = 3.9 \times 10^{14} \  \alpha^{-1/19}  \left(\frac{\textit{f}}{0.5}\right)^{-1/19} \left(\frac{D}{Mpc}\right)^{18/19} \left(\frac{S_{\nu_{op}}}{mJy}\right)^{9/19} \left(\frac{\nu_{op}}{5\ GHz}\right)^{-1} \ cm},
\end{equation}

\begin{equation}
  \mathrm{B_s = 1.0 \ \alpha^{-4/19}  \left(\frac{\textit{f}}{0.5}\right)^{-4/19} \left(\frac{D}{Mpc}\right)^{-4/19} \left(\frac{S_{\nu_{op}}}{mJy}\right)^{-2/19} \left(\frac{\nu_{op}}{5\ GHz}\right) \ G},  
\end{equation}
and 
\begin{equation}
  \mathrm{A_* = 0.82 \ \alpha^{-8/19}  \left(\frac{\epsilon_B}{0.1}\right)^{-1} \left(\frac{\textit{f}}{0.5}\right)^{-8/19} \left(\frac{D}{Mpc}\right)^{-8/19} \left(\frac{S_{\nu_{op}}}{mJy}\right)^{-4/19} \left(\frac{\nu_{op}}{5\ GHz}\right)^{2} \left(\frac{t}{10\ d}\right)^2 }.
\end{equation}
Here, $R_s$ is the radius of the forward shock advancing into the circumstellar medium, $f$ the filling factor of the shocked gas that emits synchroton emission and $D$ the distance to the supernova, for which we adopt $8.4 \pm 0.6$ Mpc \citep{2012A&A...540A..93V}. Furthermore, $B_s$ is the magnetic field strength of the shocked circumstellar gas, $\epsilon_B$ the fraction of the forward shock energy that goes into magnetic field energy density, and $\alpha$ the ratio of the fraction of the forward shock energy that goes into the energy density of relativistic electrons ($\epsilon_e$) to $\epsilon_B$. $A_*$ measures the density of the circumstellar gas such that $\rho_w = 5\times10^{11} A_* r^{-2}$ g~cm$^{-3}$. $A_*$ therefore corresponds to a mass-loss rate in units of 
$10^{-7} (v_w / 10~\rm km\ s^{-1})^{-1}$~M$_\odot$~yr$^{-1}$ assuming the circumstellar gas arises from steady mass loss from the progenitor with the wind speed $v_w$. We can estimate the retardation of the circumstellar shock assuming $R_s \propto t^m$, with values taken from Table \ref{tab:fitSSA} and we find $m = 0.76\pm0.26$, which is consistent with $m = 0.87\pm0.07$ found by \citet{2012ApJ...750L..40K} for the first $\sim 100$ days of evolution. We also note that the density profile of the ejecta in the hydrodynamical model of \citet{2019ApJ...875...17K} is roughly $\rho \propto r^{-6}$ for the ejecta that are traversed by the reverse shock during the first few years. For a constant mass-loss rate characterizing the wind, $m=(n-3)/(n-2)$ ($n$ being the index of the density slope of the ejecta). This means $m \sim 0.75$ in the hydrodynamical models by \citet{2019ApJ...875...17K}. For the magnetic field strength we obtain $B_s \propto t^q$, with $q = -0.74\pm 0.16$, which means that $B_s$ is inversely proportional to $R_s$, as has been assumed to the general case in previous modeling (e.g., \citealt{2017hsn..book..875C}).

For $v_w = 1000$~km/s, we find a mass-loss rate of $(9.6\pm0.5) \times 10^{-5}$~M$_\odot$~yr$^{-1}$, and for wind velocity 20 km/s, $(1.9\pm0.1) \times 10^{-6}$~M$_\odot$~yr$^{-1}$ for the 2012 epoch. For the 2014 epoch, we find mass-loss rates of $(2.2\pm0.6) \times 10^{-4}$~M$_\odot$~yr$^{-1}$ for a wind velocity of 1000 km/s and $(4.4\pm1.2) \times 10^{-6}$~M$_\odot$~yr$^{-1}$ for wind velocity 20 km/s. These numbers are $\sim 2.7~(6)$ (for 497 and 1197 days, respectively) times higher than estimated by \cite{2012ApJ...750L..40K} for the early phase, but smaller than $A_* = 40$ by \citet{2019ApJ...875...17K}, who, however, used the lower values $\epsilon_B = 0.04$ and $\epsilon_e = 0.03$ than our $\epsilon_B = \epsilon_e = 0.1$. If we use their values for $\epsilon_B$ and $\epsilon_e$ we would get $\sim 2.8$ times larger $A_*$ values than those in Table \ref{tab:fitSSA} \edit1{(since $\dot M \propto \alpha^{-8/19} \epsilon_B^{-1}$}) and consistent with \citet{2019ApJ...875...17K}. \edit1{A caveat is that \citet{2019ApJ...875...17K} used too low observed fluxes (cf. above), which should mean that their estimate of $\dot M$ should still be higher than ours. Regardless of that,} the mass-loss rate of the progenitor \edit1{may have been} higher earlier \edit1{in the evolution of the progenitor} than just prior to the explosion \edit1{or there was a change in one, or both, of the $\epsilon$ parameters as the circumstellar shock evolved.} 

\subsubsection{SN 1994I}

SN 1994I is a Type Ib/c supernova first detected by \cite{1994IAUC.5961....1P}. For our assumption of 8.4 Mpc, following the expansion of SN 1994I derived by  \cite{2011ApJ...740...79W}, we find $\mathrm{1.38(t_{age}/1d)~\mu as}$. This gives a size of 10.3 mas for SN 1994I at the time of our observation with LOFAR (7466 days).

\cite{2007AJ....133.2559M} detected SN 1994I with the VLA almost a decade after explosion in two different frequency bands, and derive a spectral index of -1.04. Scaling the data at 1.4 GHz to 145 MHZ using optically thin synchrotron emission scaling, $\mathrm{S_{0.145}/S_{1.4}}=(0.145/1.4)^{-1.04}$, we get a flux of  1.69 mJy at 2927 days. With this scaled flux at 2927 days, and using the light curve model in \cite{2011ApJ...740...79W}, where flux density decreases with time following $\mathrm{S \propto t_{age}^\beta}$, and $\beta =$ -1.42, we estimate a flux of  0.46 mJy at the ILT observation of 7466 days. Our detection threshold is 0.23 mJy and given that the supernova was beginning to fade in the 6~cm observations a decade before our observations, it is likely that it has entered a region with less mass loss in the wind, or it could be due to a weakening of the reverse shock if it has entered a flatter density profile of the ejecta as may have been the case for SN~1993J \citep{2015ApJ...813...43B,2019ApJ...875...17K}.

\subsubsection{SN 2005cs}
\begin{figure}[ht!]
\includegraphics[width=0.85\textwidth]{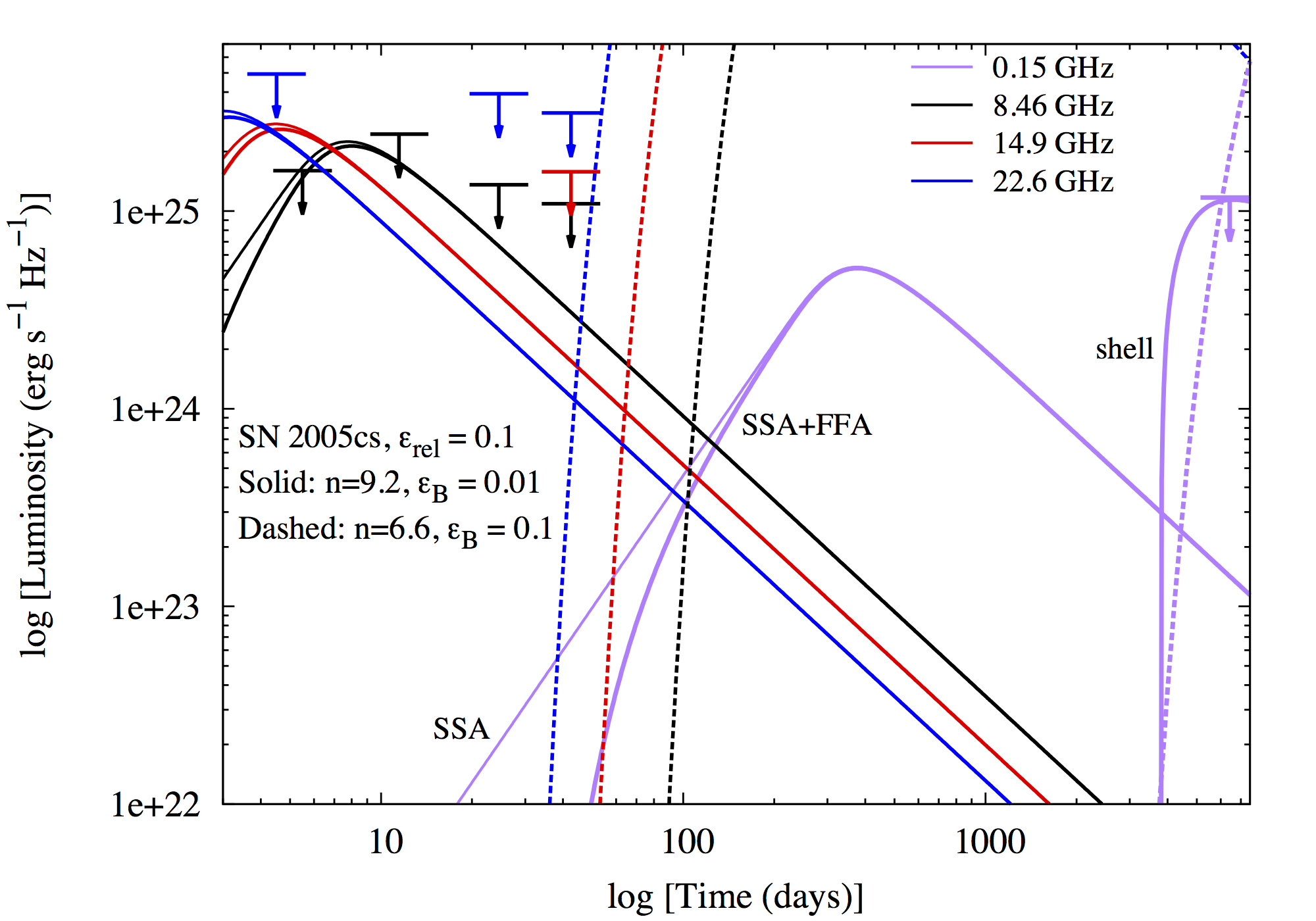}
\caption{Radio modelling of SN 2005cs. For the solid lines we have used the parameters $n=9.2$, $\epsilon_B = 0.01$, and $\epsilon_e = 0.1$. The mass-loss rate for the wind case is $M_\odot = 5\times10^{-7} (v_w / 10~\rm km/s)^{-1}$~M$_\odot$~yr$^{-1}$, whereas for the shell case it is lower inside and $\sim 3$ times larger outside $4\times10^{17}$ cm. The dashed lines are for a model with $n=6.6$, $\epsilon_B = 0.1$, $\epsilon_e = 0.1$, and $M_\odot = 8.5\times10^{-5} (v_w / 10~\rm km/s)^{-1}$~M$_\odot$~yr$^{-1}$. See text for further details and Table \ref{tab:2005csLog} for the data in the plot. ILT data from this work is labelled as 0.15 GHz. \label{fig:05cs}}
\end{figure}

Discovered by \cite{2005IAUC.8553....1K} on June 28.9, SN 2005cs was found to be a relatively underluminous Type II plateau supernova (Type~IIP), classified by \citet{2006MNRAS.370.1752P} as low-luminosity, or LL Type IIP. The supernova was  observed in the radio shortly after the explosion, but was not detected \citep{2005IAUC.8603....2S}. We have included those 3$\sigma$ upper limits, as well as our upper limit from the ILT image, in Table \ref{tab:2005csLog} and Fig. \ref{fig:05cs}. 

A recent attempt by \cite{2022MNRAS.514.4173K} to model the optical light curves for SN 2005cs, as well as another LL Type IIP, namely SN 2020cxd, showed an explosion energy of $E = 7\times10^{49}$ erg, and an ejecta mass of $M = 7.4~{\rm M}_\odot$ (their model s9.0 provided good fits). The explosion simulation was made in 3D and indicated strong asymmetry of the ejecta, but \cite{2022MNRAS.514.4173K} also made spherically symmetric averages of s9.0 which we have used as a guide for our modelling of the radio data. We simplify the ejecta structure further by assuming that the ejecta are homologous and consist of an inner structure with a density profile $\rho_i \propto V^{-a}$ and an outer structure with $\rho_o \propto V^{-n}$. The break in density structure then occurs at the velocity 

\begin{equation}
	V_{\rm b}=\sqrt {\frac{2 E}{M} \frac{\left(n-5\right) \left(5-a\right)} {\left(n-3\right)\left(3-a\right)}},
\label{eq:Luminosity}
\end{equation}
which for parameters of the s9.0 model results in a break between $(0.98-1.25)\times10^3$~km~s$^{-1}$ for  $a \in [0,1]$ and $n \in [8,14]$. 
In the angle-averaged s9.0 model a break in velocity at least ~800~km~s$^{-1}$ occurs, and $a \sim 0$, close to 3 days after explosion. These velocities agree with the photospheric velocity for s9.0 around $100-120$ days, which is also just before the observed lightcurve drop for SN~2005cs.
To set the transition between the outermost ejecta and circumstellar medium, we assume that the progenitor had constant $\dot M / v_w$, where $\dot M$ is the mass-loss rate and $v_w$ the wind speed. This gives a density dependence for the circumstellar medium which is $\rho_w \propto r^{-2}$.
We further use the information that the maximum ejecta velocity at early epochs may be as high as $1.2\times10^4$~km~s$^{-1}$ \citep{2006MNRAS.370.1752P}.
The ratio $\dot M / v_w$ is then found from setting $\rho_o = \rho_w$ at the radius $3.1\times10^{14}$ cm which is how far the outermost ejecta have reached at 3 days. We find that these requirements indicate that 
$\dot M < 6.0 \times 10^{-6} (v_w / 10~\rm km/s)^{-1}$~M$_\odot$~yr$^{-1}$ for $n \geq 8$ and $a \in [0,1]$.

To limit the likely value of $n$ further, we use the estimated electron scattering optical depth $\tau_e \sim 2/3$ through the outermost ejecta. 
\cite{2022MNRAS.514.4173K} estimates that the velocity at which this occurs is positioned close to the photosphere for the first $\sim 20$ days. 
At 20 days, this ejecta velocity is $\sim 2000$~km~s$^{-1}$. In our models, assuming fully ionized ejecta consisting of hydrogen and helium with a ratio of He/H = 0.1, $\tau_e (\rm 20~days) \geq 1$ through ejecta with velocities $> 2000$~km~s$^{-1}$, for $a \in [0,1]$ and $n < 15$. 
However, an electron fraction of $\sim 0.2$ may be more likely (e.g., \citealt{2022A&A...666A.104E}), which in our case limits $n$ to the range $n \in [9,14]$, but to allow for even lower electron fractions, we also include $n=8$. Guided by these estimates, and since models with $a=0$ and $a=1$ do not differ to any larger extent, we will concentrate on models with $n \in [8,14]$ and a=0. The mass-loss rates for these $n$-values are $\dot M  \approx 5~(0.6)\times10^{-6} (v_w / 10~\rm km/s)^{-1}$~M$_\odot$~yr$^{-1}$ for $n = 8~(9)$, and  $\dot M < 10^{-7} (v_w / 10~\rm km/s)^{-1}$~M$_\odot$~yr$^{-1}$ for $n \geq 10$. 

To model the radio emission from the interaction between the ejecta and the circumstellar medium we use the similarity solutions and methods of \cite{1982ApJ...258..790C} and \cite{2017hsn..book..875C}. The propagation of the radius of the interaction region $R_s \propto t^{(n-3)/(n-2)}$, and we assume that the fraction $\epsilon_B$ of the forward shock energy density $\rho_w V_{\rm s}^2$, goes into magnetic field energy density, and that the fraction $\epsilon_e$ goes into relativistic electron energy density. For the relativistic electrons, we assume a power law distribution of the electron energies, $dN/dE = N_0E^{-p}$, where $E=\gamma m_ec^2$ is the energy of the electrons and $\gamma$ is the Lorentz factor.
The intensity of optically thin synchrotron emission is then $\propto \nu^{-\alpha}$, where $\alpha = (p-1)/2$. We have used $p=3$. The volume of the region that generates synchrotron emission is assumed to be the entire volume between the forward shock and the contact discontinuity between \edit1{shocked} ejecta and shocked circumstellar gas. The synchrotron emission suffers both from SSA and external free-free absorption (FFA). A crucial parameter for FFA is the temperature of the circumstellar gas, which is uncertain (e.g., \citealt{1988A&A...192..221L}). We have used $5\times10^4$~K.
We have also assumed that H and He in the circumstellar gas to be fully ionized, which may overestimate the degree of ionization for a LL Type IIP.
With these assumptions, we show in Figure \ref{fig:05cs} the modeled radio emission (with and without FFA) for a model with $n=9.2$, $\epsilon_B = 0.01$ and $\epsilon_e = 0.1$. The mass-loss rate is $5\times10^{-7} (v_w / 10~\rm km/s)^{-1}$~M$_\odot$~yr$^{-1}$. For this, and smaller values 
of $\dot M / v_w$, modeled radio fluxes are lower than the observed upper limits, and FFA is less important than SSA at the peak fluxes of each observed frequency. It is also obvious that at the epoch of the LOFAR observations, the modeled flux is about two orders of magnitude below detection limit. In the $n=9.2$ model, the maximum velocity of unshocked ejecta at this epoch is $\sim 4060$~km~s$^{-1}$, which is too slow to generate any appreciable synchrotron emission. It is also possible to avoid radio detection for very dense winds since FFA will then operate efficiently and make radio emission non-detectable. To estimate the lowest mass-loss rate needed for this to happen, we have run models with $\epsilon_B = \epsilon_e = 0.1$ and $a=0$, and find that no emission would be detected for models with $n \leq 6.6$, which all have
$\dot M \geq 8.5\times10^{-5} (v_w / 10~\rm km/s)^{-1}$~M$_\odot$~yr$^{-1}$, with the 150 MHz observations being decisive. We have included 
a model with $n=6.6$ and $\dot M = 8.5\times10^{-5} (v_w / 10~\rm km/s)^{-1}$~M$_\odot$~yr$^{-1}$ in Figure \ref{fig:05cs}. 
We find this scenario to avoid radio detection less likely since $\tau_e \geq 50$ (for fully ionized ejecta) at 20 days in these models, 
which would require an electron fraction in the ejecta at 20 days which is less than $\sim 0.01$ to be compatible with the results of \cite{2022MNRAS.514.4173K}. \edit1{Moreover, such a dense wind may produce narrow circumstellar spectral lines which are not detected, hence eliminating such a scenario.} If we had assumed $\epsilon_B = 0.01$ as for the SSA dominated models in Figure \ref{fig:05cs}, the wind density would have been even more extreme.
         
To obtain a detectable 150 MHz signal at 6428 days without producing detected emission at higher frequencies during early epochs, the circumstellar 
medium could be dilute close to the supernova, and denser at larger radii. Figure \ref{fig:05cs}  includes the 150 MHz lightcurve for such a circumstellar 
``shell" model. Here we have assumed that the maximum ejecta speed is $\sim 9600$~km~s$^{-1}$ at $\sim 3900$ days when the circumstellar shock runs into a shell characterized by the progenitor mass-loss rate $\dot M \approx 1.8\times10^{-6} (v_w / 10~\rm km/s)^{-1}$~M$_\odot$~yr$^{-1}$. (This means that we have assumed almost no deceleration of the outermost ejecta until they enter the region of increased circumstellar density.) Parameters are otherwise the same as for the other $n=9.2$ model in Figure \ref{fig:05cs} (i.e., $n = 9.2$, $\epsilon_B=0.01$ and $\epsilon_e=0.1$). The inner radius of increased circumstellar density is $4\times10^{17}$~cm.
Since it takes a few expansion time scales to set up a similarity solution, we have scaled the flux with the swept-up mass since the \edit1{shock} crossed $4\times10^{17}$~cm compared to the swept-up mass for a ``mature'' similarity solution. The swept-up mass scales linearly with radius for a $\rho_w \propto r^{-2}$ wind, so the scaling factor is simply $1 - (4\times10^{17}/R_s)$. 
\citep[For a more detailed discussion on shell interaction and how quickly a similarity solution is obtained, see][]{2016ApJ...823..100H}. At 6428 days the circumstellar shock is at $6.2\times10^{17}$~cm, and the swept-up mass of the high-density region is $\sim 0.013$~M$_\odot$. This gives a rough estimate of how the LOFAR observation constrains a possible high-density shell at these radii.

\edit1{The results in Figure \ref{fig:05cs} were derived for $\epsilon_e =0.1$ and $\epsilon_B = [0.01,0.1]$. These are values often used to estimate $\dot M / v_w$ for radio SNe \citep[e.g.,][]{2020ApJ...890..159L}, and within this range of values for $\epsilon_B$, we used $\epsilon_B = 0.01$ to obtain the highest possible upper limit of $\dot M / v_w$ in the SSA-dominated case. For the FFA-dominated case, we used to get the lowest possible lower limit of $\dot M / v_w$. However, we emphasize that there is considerable uncertainty in the values for $\epsilon_B$ and $\epsilon_e$. In a recent investigation \citet{2021ApJ...917...55R} found that $0.001 \lesssim \epsilon_B \lesssim 0.1$ and $10^{-4} \lesssim \epsilon_e \lesssim 0.05$ for six young SN remnants, and that there are variations of $\epsilon_B$ and $\epsilon_e$ even within the same remnant. If we include this information it will not affect the lowest possible lower limit in the FFA-dominated case for SN 2005cs in Figure \ref{fig:05cs} (since this was estimated from large values of $\epsilon_e$ and $\epsilon_B$), but it will affect the highest possible upper limit of $\dot M / v_w$ in the SSA-dominated case. If we fix $\epsilon_B$ at $\epsilon_B = 0.1$, and set $\epsilon_e = 0.05~(0.001,2\times10^{-4})$, then the upper limit of $\dot M$ becomes $\dot M \lesssim 2~(17,390)\times10^{-7} (v_w / 10~\rm km/s)^{-1}$~M$_\odot$~yr$^{-1}$, for $n=9.5~(8.5,7.0)$ and $a=0$. For $\epsilon_e = 10^{-4}$ there is no solution for $n\geq6$. However, if we only allow $n\geq8$ (cf. above), the smallest possible value for $\epsilon_e$ is $3\times10^{-4}$, for which $\dot M \lesssim 5\times10^{-6} (v_w / 10~\rm km/s)^{-1}$~M$_\odot$~yr$^{-1}$.
We can also fix $\epsilon_e$ at $\epsilon_e = 0.05$ and set $\epsilon_B = 0.1~(0.01,0.001)$. We then find $\dot M \lesssim 2~(7,26)\times10^{-7} (v_w / 10~\rm km/s)^{-1}$~M$_\odot$~yr$^{-1}$, for $n=9.5~(8.9,8.3)$ and $a=0$. In all models with $\dot M \gtrsim 10^{-6} (v_w / 10~\rm km/s)^{-1}$~M$_\odot$~yr$^{-1}$, FFA starts to become important also for the `SSA-dominated' case.}

\edit1{In summary, due to the unknown values of $\epsilon_e$ and $\epsilon_B$, there is considerable uncertainty of the the highest possible upper limit of $\dot M / v_w$ in the SSA-dominated case. However, looking at other SNe IIP, there are hints of deviation from energy equipartition between relativistic electrons and magnetic field strength with $\alpha \sim 10-100$, and the derived mass-loss rates are in the range $\dot M \sim (1-10)\times10^{-6} (v_w / 10~\rm km/s)^{-1}$~M$_\odot$~yr$^{-1}$ \citep[e.g.,][and referenced therein]{2022A&A...666A..82R}. If we assume $\epsilon_e = 0.05$ and $\alpha =30$, i.e., $\epsilon_B = 0.0017$ for SN~2005cs, we obtain an upper limit of $\dot M \lesssim 1.7\times10^{-6} (v_w / 10~\rm km/s)^{-1}$~M$_\odot$~yr$^{-1}$ (for $n=8.5$ and $a=0$), which could argue for the progenitor of SN~2005cs having somewhat less mass-loss than for the average SN IIP, unless $\epsilon_e = 0.05$ and $\alpha \gtrsim 100$. For the same reasons, the limit on $\dot M$ characterizing a high-density shell probed by the LOFAR data is higher if we allow for lower values of $\epsilon_e$ and $\epsilon_B$. In this case, the limit for such a shell is instead $\dot M \lesssim 6\times10^{-6} (v_w / 10~\rm km/s)^{-1}$~M$_\odot$~yr$^{-1}$ (for $n=8.5$, $a=0$, $\epsilon_e = 0.05$, and $\alpha =30$.}

\begin{table*}
\caption{Parameters of SNe 1994I and 2005cs and 2011dh}
\scalebox{0.9}{
\begin{tabular}{llccccc}
\tableline\tableline
Supernova  & Date of observation & Time after explosion  & Central Frequency &  Flux Density    & Luminosity & Reference \cr
                    &     (UT)                     &  (days)                      &       (GHz)   &  ($\mu$Jy)  & ($10^{25}$~erg~s$^{-1}$~Hz$^{-1}$)    &              \cr
\tableline

SN 1994I           &  2004 Jan 5           &  3567.00                         & 1.425        & $160\pm22$            &  $1.35\pm0.19$         & 1 \cr
                   &  2004 Jan 5           &  3567.00                         & 4.860        & $46\pm11$            & $0.39\pm0.09$          & 1 \cr
                   &  2011 Nov 7           &  6428.00                  & 1.65         & $<50$                 &  $<0.42$  & 2    \cr
                   &  2014 Sep 10         &  7466.00                  & 0.145         & $<138$                   &  $<1.17$  & 3    \cr
    &  &   &         &       &    &       \cr
SN 2005cs  &  2005 Jul 2.01          & 4.51                          & 22.64           &   $<585$    &   $<4.94$   &   4         \cr
                    &  2005 Jul 3.00         &  5.50                          & 8.460           &   $<189$    &   $<1.60$  &   4         \cr  
                    &                                &                                    & 14.94         &   $<900$    &   $<7.60$   &   4         \cr  
                     &  2005 Jul 8.97         &  11.47                        & 8.460           &   $<291$    &   $<2.46$   &   4         \cr  
                    &                                &                                    & 22.64         &   $<918$    &   $<7.75$   &   4         \cr 
                    &  2005 Jul 22.04       &  24.54                      & 8.460           &   $<161$    &   $<1.36$   &   4         \cr  
                    &                                &                                    & 22.64         &   $<464$    &  $<3.92$    &   4         \cr
                    &  2005 Aug 9.00       &   42.50                      & 8.460           &   $<129$    &    $<1.09$  &   4         \cr  
                    &                                 &                                 & 14,94          &  $<187$  &   $<1.58$   &   4         \cr 
                    &                                &                                    & 22.64         &   $<372$    &  $<3.14$    &   4         \cr 
                    &  2014 Sep 10         &  3361.5                  & 0.145         & $<138$                   &  $<1.17$  & 3    \cr
&  &   &         &       &    &       \cr
           
SN 2011dh  &  2012 July 20  &  415  & 0.323   &  $3610\pm250$  & $30.5\pm2.1$  & 5 \cr
           &  2012 July 21  &  416  & 0.607   &  $4240\pm70$   & $35.8\pm0.6$  & 5 \cr
           &  2012 July 26  &  421  & 1.387   &  $3600\pm50$  & $30.4\pm0.4$  & 5 \cr
           
           &  2012 Aug 1   &  427    &  8.4     &  $880\pm60$  & $7.4\pm0.5$  & 3 \cr 
                          &   &        &  8.55    &  $778\pm24$& $6.6\pm0.2$  & 3 \cr
                           &   &            &  9.56    &  $663\pm23$& $5.6\pm0.2$  & 3 \cr
                          &   &             &  13.5    &  $507\pm17$& $4.3\pm0.1$  &  3 \cr
                         &   &              &  14.5      &  $466\pm17$& $3.9\pm0.1$  & 3 \cr
           
           & 2014 Jan 31   &  975  &  4.615  &  $1099\pm43$  & $9.2\pm0.4$  & 3 \cr
           &    &   &  4.871  &  $949\pm29$  & $8.0\pm0.2$  &  3\cr
           &    &   &  5.127  &  $925\pm30$  &  $7.8\pm0.2$ &  3\cr
           &    &    &  5.383  &  $926\pm42$  & $7.8\pm0.3$  &  3\cr
           &  &    &  6.715  &  $731\pm34$  & $6.2\pm0.3$  &  3\cr
           &    &   &  6.971  &  $754\pm26$  & $6.4\pm0.2$  &  3\cr
           &    &   &  7.227  &  $679\pm29$  & $5.7\pm0.2$  &  3\cr
           &    &    &  7.483  &  $731\pm46$  & $6.2\pm0.4$  &  3\cr

            &  2014 Sep 10  &  1197   &  0.145   &  $5200\pm900$    &  $43.9\pm7.6$ & 6 \cr  
                             & &       &  0.145   &  $970\pm140$  & $8.19\pm1.18$  & 3 \cr
           
           &  2014 Oct 18  &  1235  & 4.415   & $866\pm106$   & $7.3\pm0.9$  & 3 \cr
           &    &    & 4.671   & $869\pm83$   & $7.3\pm0.7$  &  3\cr
           &     &    & 4.927   & $762\pm74$   &  $6.4\pm0.6$ &  3\cr
           &     &    & 5.183   & $703\pm73$   & $5.9\pm0.6$  &  3\cr
           &     &    & 7.015   & $507\pm54$   & $4.3\pm0.4$  &  3\cr
           &    &    & 7.271   & $524\pm48$   & $4.4\pm0.4$  &  3\cr
           &    &    & 7.527   & $703\pm51$   & $5.9\pm0.4$  &  3\cr
           &    &    & 7.783   & $515\pm50$   & $4.3\pm0.4$  &  3\cr
           
           &    &    &  19.315  & $130\pm20$   & $1.1\pm0.2$  & 3 \cr
           &    &   &  21.997  & $116\pm25$   & $9.8\pm0.2$  & 3 \cr
           &    &    &  24.681  & $94\pm23$   & $0.8\pm0.2$  &  3\cr

           &  2016 Jan 8  &  1682  &  0.325  & $5200\pm500$   & $43.9\pm4.2$  &  7\cr
           
\tableline
\end{tabular}
}
\tablecomments{The columns starting from left to right are as follows: Supernova name; Date of observation;  Time after explosion, where UT 2005 June 27.5 has been used as the explosion date for SN 2005cs \citep{2006MNRAS.370.1752P} and UT  1994 Mar 31 has been used as the explosion date for SN 1994I \citep{2011ApJ...740...79W}; Central frequency of observation; flux densities and $3\sigma$ upper limit on the same; luminosity and $3\sigma$ upper limit on the same: (1) \citet{2007AJ....133.2559M} (2) \citet{2015MNRAS.452...32R}, (3) This work, (4) \citet{2005IAUC.8603....2S}, (5) \citet{2016MNRAS.459..595Y}, (6) LoTSS catalogue at \SI{6}{\arcsecond} resolution,  (7) private communication, Subhashish Roy. 
}
 \label{tab:2005csLog}
\end{table*}

\subsection{M 51a AGN related sources}
\begin{figure}[ht!]
\includegraphics[width=\textwidth]{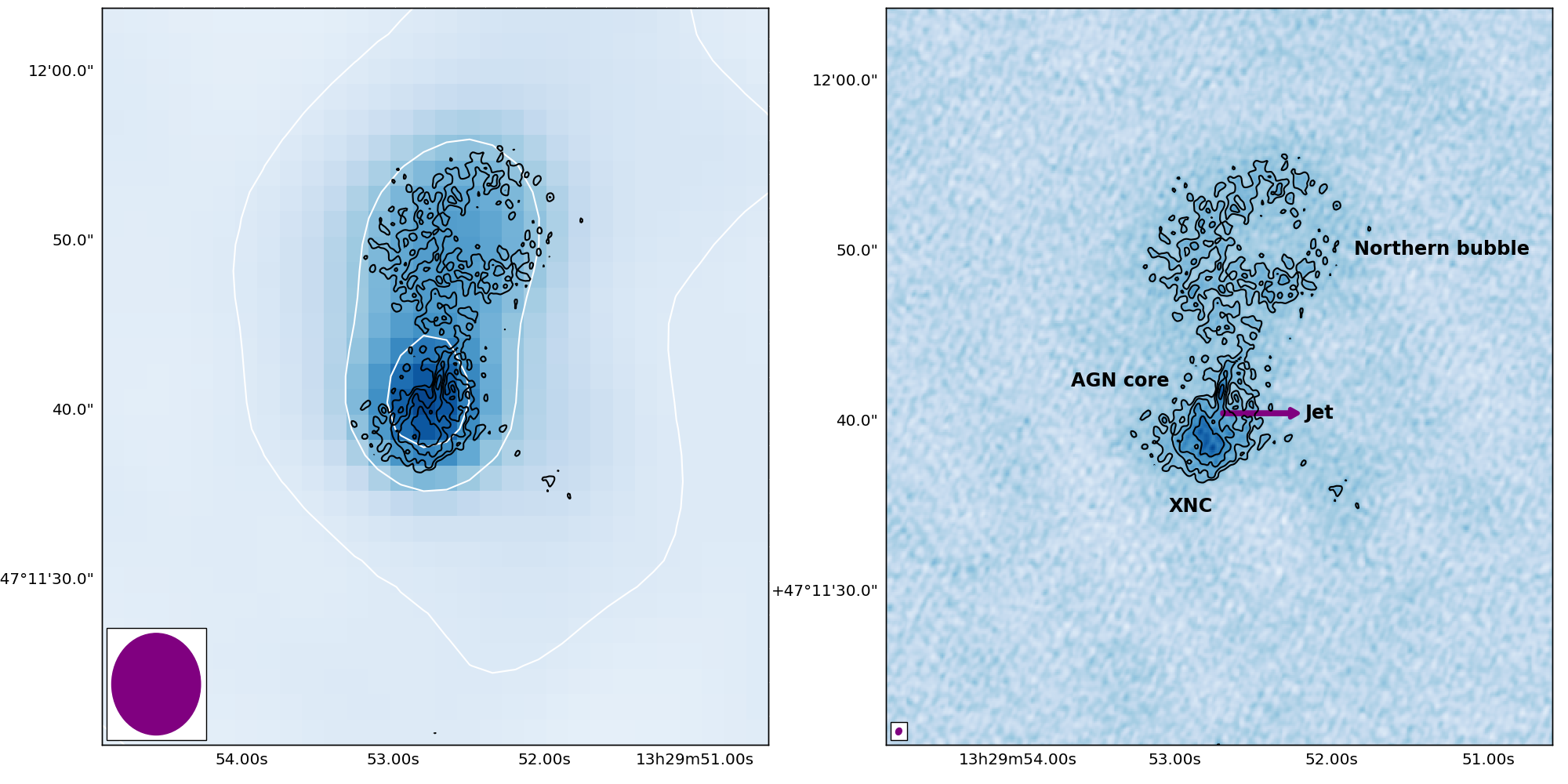}
\caption{ Left: ILT image of the centre of M 51 in black contours overlaid on LoTSS \SI{6}{\arcsecond} resolution image in the background and also in white contours. Contours from 2 $\sigma$ to 5 $\sigma$ for the ILT, and contours from 1 $\sigma$ to 5 $\sigma$ for LoTSS. Right: ILT image in background and also in black contours overlaid. Linear stretch from -5 to +5 $\sigma$. The restoring beam for the two LOFAR images are
shown in the bottom left corner.  \label{fig:M 51centre}}
\end{figure}

The morphology of the radio nuclear emission from the centre of M 51 in the ILT image (Fig. \ref{fig:M 51centre}) is similar to the morphology seen in other radio studies by, eg., \cite{2011AJ....141...41D} and \cite{2007AJ....133.2559M} using the VLA, with the AGN core in the centre, the extranuclear cloud (XNC) to the south and ring-like emission from the Northern bubble above. However, while the VLA observations only show a compact source at the center, the radio morphology seen in the ILT image is a narrow, elongated structure towards the south like the EVN image in \cite{2015MNRAS.452...32R}, albeit without resolving the sub-components in the AGN core.  The position of the AGN  (Table \ref{tab:sources}), agrees well with the position reported by \cite{2015MNRAS.452...32R}. Since the AGN core is unresolved, we cannot describe the sub-components in detail as described by \cite{2015MNRAS.452...32R}, but the \textsc{PyBDSF} fit for the AGN core is given in Table \ref{tab:sources} (Source 1).

The XNC morphology is similar to the one described by \cite{1992AJ....103.1146C}, with a thin jet connecting the AGN to the XNC. They also describe a bow shock in the XNC from the interaction of the jet with the surrounding Interstellar Medium, which seems likely from the ILT image as well. \cite{2004ApJ...603..463B} and \cite{1985ApJ...293..132F} also suggest continuous fuelling of the XNC by the jet. We estimate a flux density of 71.45 $\pm$ 3.00 mJy for the XNC using the task IMSTAT CASA.

The ILT image also shows the northern bubble that is thought to be blown out due to an earlier ejection cycle from the AGN \citep{2007AJ....133.2559M}. It is similar to the structure reported by them indicating cooler gas within. For the northern bubble, since \textsc{PyBDSF} does not pick up emission from the entire region, we estimate flux density of 131.44 $\pm$ 0.12 mJy using the task IMSTAT in CASA. 

In addition, we also detect extended emission coincident with the position of Component N. It is possible that our image at 145 MHz has almost resolved out emission from Component N and hence only the faint detection of extended emission at this resolution rather than that of a compact source. We also convolved the image with a bigger beam of \SI{1.08}{\arcsecond} $\times$ \SI{1.02}{\arcsecond} (similar to VLA-A 20cm image in \citealt{2015MNRAS.452...32R}) and compared it with the \SI{6}{\arcsecond} LoTSS image. Even when convolved with a bigger beam, the source remains extended and is spread over an area of about \SI{15}{\arcsecond}. We estimate a flux density of 55.08 $\pm$ 8.08 mJy using IMSTAT in CASA (this flux density matches what is estimated by \textsc{PyBDSF} at the same position with a source size of \SI{34}{\arcsecond} $\times$ \SI{17}{\arcsecond}). Given that we do not detect a compact source, we do not include Component N in the source list.  \cite{2015MNRAS.452...32R} suggest a study of the synchrotron aging of Component N, as discussed in \cite{1991ApJ...378...65C}. However, we see no evidence of a break frequency in the VLA data in literature, and therefore we do not see evidence of synchrotron aging in component N. Also,  if the emission  we detect at the position of component N comes truly from it, as it is most likely the case, then there is no absorption at LOFAR frequencies. We therefore conclude that Component N may indeed be a fossil radio hotspot as suggested by \cite{2015MNRAS.452...32R}, that has left behind a diffuse shell from the loss of continuous energy from the AGN. 

\subsection{Other sources and their identifications}

First, a note on detecting H {\sc ii} regions: from equation 5 in \cite{1991ApJ...378...65C} and using the  M 51 ILT image, for a beam size of $\theta_M = \SI{0.436}{\arcsecond}$, $\theta_m =$\SI{0.366}{\arcsecond} and a flux density detection limit of 5 $\times$ 0.046 mJy, the lower limit of the brightness temperature for compact sources detected at 145 MHz is $10^{4.9}$K. Free-free emission from H {\sc ii} regions have brightness temperatures up to 2 $\times$ $10^4$K \citep{2009tra..book.....W}. Hence as first noted in \cite{2015A&A...574A.114V}, the resolution of the ILT image rules out that any of the detected objects are H {\sc ii} regions.
Hence, ruling out H {\sc ii} regions, candidates for the sources detected in this ILT image of M 51 are expected to be background sources, AGN, high mass X-ray binaries and supernova remnants.

\subsubsection{J133005+471035}
Source 3 in Table \ref{tab:sources}, J133005+471035 is resolved into 4 components with the EVN and compact with the VLA \citep{2015MNRAS.452...32R}. With the ILT, it is not fully resolved, but it clearly is not a single compact structure either. \citet{2015MNRAS.452...32R} conclude that a two-sided jet from an active nucleus is a possible scenario from their EVN image.

\subsubsection{[MZF2015] J1329+4717}
Source 8 in Table \ref{tab:sources} is a bright radio source visible in both LoTSS (compact) and ILT (extended emission with apparent two lobes). VLA data from \cite{2015ApJ...800...92M} classify it as an extragalactic background radio source.

\subsubsection{M 51b}
Source 9 in Table \ref{tab:sources} is the interacting companion to M 51a. A bright compact source in LoTSS, M 51b shows extended resolved emission with the ILT.

\subsubsection{J133016+471024}
Source 10 in Table \ref{tab:sources}, J133016+471024 is a single bright source in LoTSS, whereas with the ILT, two lobes of extended emission are visible, but not a central engine. (Hence there was no cross-match with LoTSS for this particular source). MERLIN and EVN images in \cite{2015MNRAS.452...32R} detect the unresolved central source while the 20 cm VLA image shows a central source with two radio lobes with its morphology and luminosity placing it as a radio-loud AGN, an FR II galaxy.

\subsubsection{CXOU J132954.9+470922}
Source 11 in Table \ref{tab:sources} is a High Mass X-ray Binary. \edit1{The source is not detected by LoTSS but has a high flux density similar to SN 2011dh and can be cross-matched with the NASA/IPAC Extragalactic Database.} \cite{2007AJ....133.2559M} find a spectral index of -0.3 for this source (listed as 65 in Table 2 of their work). From their 20cm flux density, this can be extrapolated to 145 MHz to be 817 $\mu$Jy, which matches very well with our finding of 830 $\mu$Jy from \textsc{PyBDSF}.

\section{Conclusion} \label{sec:conc}

This paper presents the first sub-arcsecond resolution ILT image of the nearby galaxy M 51 at 145 MHz. We discuss the compact sources in the galaxy, the centre of M 51, and the supernovae in the galaxy. We detect SN 2011dh with the ILT, which is also seen in the LoTSS catalogue.  Importantly, we find that the LoTSS flux density for the supernova is about five times higher than the flux density with our ILT image, which would affect the absorption scenario for the supernova, and consequently, mass-loss rates and other parameters. Using VLA and GMRT data at two different epochs, combined with the ILT flux density, we derive the radii of the forward shock, magnetic field strengths of the shocked CSM, and corresponding mass-loss rates for the progenitor at these two epochs. The time evolution of the shock characterised by the parameter \textit{m} seems consistent with previous studies. Our \edit1{analysis also indicates} that the mass-loss rate of the progenitor was higher just prior to the explosion \edit1{or that the values for one, or both of the $\epsilon$ parameters changed as the circumstellar shock evolved}. 

We do not detect SN 1994I or SN 2005cs, but we discuss both \edit1{at} some length, the former from extrapolated expected flux densities, and the latter for the early-time light curves in light of a new explosion model. For SN 1994I, we find that the upper limit we obtain on the flux density at 7466 days with the ILT could be inconsistent with a scenario of constant mass-loss rate or could indicate weakening of the reverse shock. For SN 2005cs, we find that along with the upper limits on flux density at late-times from the ILT image, the early radio data for the supernova is consistent with the explosion model presented in \cite{2022MNRAS.514.4173K} and an upper limit on the progenitor mass-loss of \edit1{$\dot M \lesssim \rm few \times10^{-6} (v_w / 10~\rm km/s)^{-1}$~M$_\odot$~yr$^{-1}$}, with no possibility of the ejecta running into a shell. This is constrained by the lack of any detectable signal at 6428 days with the ILT for which the ejecta would have to run into a high-density shell characterized by the progenitor mass-loss rate \edit1{$\dot M \gtrsim 10^{-5} (v_w / 10~\rm km/s)^{-1}$~M$_\odot$~yr$^{-1}$ to be detected}. 

We find that the morphology of the nuclear emission in  M 51 with the ILT is very similar to the morphology seen in the higher frequency VLA by other studies. With the ILT we are unable to fully resolve the AGN core, although it is not a compact source but narrow and elongated towards the XNC in the south. We also discuss the emission from the XNC (with a thin jet connecting the AGN and the XNC signalling continuous fuelling of the XNC) and the northern bubble (with the same bubble structure that is visible at higher frequencies denoting cooler gas inside and the possibility of it being left over from a previous ejection from the AGN), along with a possible detection of Component N (which we conclude may indeed be a fossil hotspot, with no evidence we can see of synchrotron aging at higher radio frequencies). 
We also present a few other interesting sources detected in the field with the ILT.

The results presented in this paper provide a proof of concept of the advantages of using the ILT to detect and characterize low frequency radio emission from compact sources (including radio supernovae) in nearby galaxies rather than relying on low \SI{6}{\arcsecond} resolution LOTSS survey data using the Dutch baselines of LOFAR to study these objects. Going to higher angular resolution removes almost completely the effects of diffuse radio emission from the host galaxy allowing more accurate flux density measures, including accurate upper limits, to be made for such compact sources. In addition our observations revealed  two new compact sources  in M 51 which was not detected in low resolution LoTSS images because it was too weak to be separated from the  diffuse emission of the galaxy. It is interesting to note that if  M51 was placed at even a slightly larger distance, or the compact sources were slightly weaker in luminosity,  then more compact sources, including SN 2011dh, would not have been detectable in the LoTSS 6’’ resolution images due to confusion with diffuse galactic continuum. In contrast these sources would still have been detectable using  the ILT. This observation  demonstrates the potential advantages  for extragalactic compact source population studies of large scale automated ILT processing towards nearby galaxies using the extensive archive of the long baseline that exists as part of the LOFAR LoTSS sky survey.

\section{Software and third party data repository citations} \label{sec:cite}

\begin{acknowledgments}
This paper is based on data obtained with the International LOFAR Telescope (ILT) under project code LC2\_038. 
LOFAR data products were provided by the LOFAR Surveys Key Science project (LSKSP; https://lofar-surveys.org/) and were derived from observations with the International LOFAR Telescope (ILT). LOFAR \citep{2013A&A...556A...2V} is the Low Frequency Array designed and constructed by ASTRON. It has observing, data processing, and data storage facilities in several countries, which are owned by various parties (each with their own funding sources), and which are collectively operated by the ILT foundation under a joint scientific policy. The efforts of the LSKSP have benefited from funding from the European Research Council, NOVA, NWO, CNRS-INSU, the SURF Co-operative, the UK Science and Technology Funding Council and the Jülich Supercomputing Centre.
LoTSS Data Release 2- The data are described by \cite{2022A&A...659A...1S}.
This research has made use of the NASA/IPAC Extragalactic Database (NED), which is funded by the National Aeronautics and Space Administration and operated by the California Institute of Technology.
This work made use of published GMRT data. We thank the staff of the GMRT that made these observations possible. GMRT is run by the National Centre for Radio Astrophysics of the Tata Institute of Fundamental Research.
P.L. acknowledges support from the Swedish Research Council. D.V. acknowledges support from Onsala Space Observatory for the provisioning of its facilities support. The Onsala Space Observatory national research infrastructure is funded through Swedish Research Council grant No 2017-00648.
MPT and JM acknowledge financial support through grants CEX2021-001131-S and PID2020-117404GB-C21 funded by the Spanish MCIN/AEI/10.13039/501100011033.
\end{acknowledgments}

\vspace{5mm}
\facilities{ILT, VLA, GMRT}

\software{\textsc{astropy} \citep{2022ApJ...935..167A},  
          \textsc{PyBDSF} \citep{2015ascl.soft02007M},
          \textsc{emcee} \citep{2013PASP..125..306F},
          \textsc{NumPy} \citep{harris2020array},
          \textsc{SciPy} \citep{2020SciPy-NMeth},
          \textsc{Matplotlib} \citep{Hunter:2007},
          \textsc{uncertainties} \citep{Lebigot:2010},
          CASA \citep{2022PASP..134k4501C},
          \textsc{WSClean} \citep{2014MNRAS.444..606O},
          LOFAR-VLBI pipeline \citep{2022A&A...658A...1M}
          }

\newpage
\bibliography{sample631}{}
\bibliographystyle{aasjournal}

\listofchanges
\end{document}